\input harvmac

\font\ensX=msbm10
\font\ensVII=msbm7
\font\ensV=msbm5
\newfam\math
\textfont\math=\ensX \scriptfont\math=\ensVII \scriptscriptfont\math=\ensV
\def\ensemble{\fam\math\ensX}
%%%%%%%%%%%%%%%%%%%%%%%%%%%%%%%%%%%%%%%%%%%%%%%%%%%%%%%%%%%%%%%%%%%%%%%%%%%%
\overfullrule=0pt
%\draftmode
%%%%%%%%%%%%%%%%%%%%%%%%%%%%%%%%%%%%%%%%%%%%%%%%%%%%%%%%%%%%%%%%%%%%%%%%%%%%%%
\def\ZZ{{\ensemble Z}}
\def\RR{{\ensemble R}}
\def\II{{\ensemble I}}
\def\tfrac#1#2{{#1\over #2}}
\def\teight#1{t_{(8)}^{(#1)}}
\def\thuit#1{t_{(8)}{}^{{#1}_1\cdots {#1}_8}}
\def\ehuit#1{\varepsilon_{(8)}{}^{{#1}_1\cdots {#1}_8}}
\def\tria{$\triangleright$}
\def\slashed#1{\not\!\!#1}
%%%%%%%%%%%%%%%%%%%%%%%%%%%%%%%%%%%%%%%%%%%%%%%%%%%%%%%%%%%%%%%%%%%%%%%%%%%%%%
\lref\sen{A.~Sen, {\sl Descent relations among bosonic D-branes}, Int.\ J.\
Mod. Phys. A {\bf 14} (1999) 4061, hep-th/9902105\semi
A.~Sen, {\sl Non-BPS states and branes in string theory}, hep-th/9904207\semi
%%CITATION = HEP-TH 9904207;%%
%%CITATION = HEP-TH 9902105;%%
A.~Sen, {\sl Universality of the tachyon potential}, JHEP {\bf 9912} (1999) 027, hep-th/9911116.
%%CITATION = HEP-TH 9911116;%%
}

\lref\BerkovitsBT{N.~Berkovits, {\sl A new approach to superstring field
theory}, Fortsch.\ Phys.\  {\bf 48} (2000) 31, hep-th/9912121.
%%CITATION = HEP-TH 9912121;%%
}
\lref\ShatashviliUX{S.~L.~Shatashvili, {\sl On field theory of open strings,
tachyon condensation and closed strings}, Talk at Strings 2001, Mumbai, India,
hep-th/0105076. 
%%CITATION = HEP-TH 0105076;%%
}
\lref\GrossYK{D.~J.~Gross and W.~Taylor, {\sl Split string field theory II},
hep-th/0106036.
%%CITATION = HEP-TH 0106036;%%
}
\lref\GerasimovGA{A.~A.~Gerasimov and S.~L.~Shatashvili, {\sl On exact tachyon
potential in open string field theory}, JHEP {\bf 0010} (2000) 034,
hep-th/0009103\semi
%%CITATION = HEP-TH 0009103;%%
D.~Kutasov, M.~Mari{\~n}o and G.~Moore,
{\sl Some exact results on tachyon condensation in string field theory},
JHEP {\bf 0010}  (2000) 045, hep-th/0009148\semi
%%CITATION = HEP-TH 0009148;%%
D.~Kutasov, M.~Mari{\~n}o and G.~Moore,
{\sl Remarks on tachyon condensation in superstring field theory},
hep-th/0010108\semi
%%CITATION = HEP-TH 0010108;%%
A.~A.~Gerasimov and S.~L.~Shatashvili, {\sl
Stringy Higgs mechanism and the fate of open strings}, JHEP {\bf 0101} (2001)
019, hep-th/0011009.
%%CITATION = HEP-TH 0011009;%%
}
\lref\ShatashviliKK{E.~Witten, {\sl On background independent open string field
theory}, Phys.\ Rev.\ D {\bf 46}, 5467 (1992), hep-th/9208027\semi
E.~Witten, {\sl Some computations in background independent off-shell string
theory}, Phys.\ Rev.\ D {\bf 47}, 3405 (1993), hep-th/9210065\semi
%%CITATION = HEP-TH 9208027;%%
%%CITATION = HEP-TH 9210065;%%
S.~L.~Shatashvili, {\sl Comment on the background
independent open string theory}, Phys.\ Lett.\ B {\bf 311}, 83 (1993),
hep-th/9303143\semi
S.~L.~Shatashvili, {\sl On the problems with background independence in string
theory}, Algebra and Anal., {\bf 6} (1994) 215, hep-th/9311177.
%%CITATION = HEP-TH 9311177;%%
%%CITATION = HEP-TH 9303143;%%
}
\lref\rfWittenKtheory{E. Witten, {\sl D-branes and K-theory}, JHEP, {\bf 12}
(1998) 019,  hep-th/9810188.
%%CITATION = JHEPA,9812,019;%%
}
\lref\wone{E. Witten, {\sl Interacting Field Theory of Open Superstrings},
Nucl.Phys. B{\bf 276} (86) 291.
%%CITATION = NUPHA,B276,291;%%
}
\lref\BachasMC{C.~Bachas, C.~Fabre, E.~Kiritsis, N.~A.~Obers and P.~Vanhove,
{\sl Heterotic/type-I duality and D-brane instantons},
Nucl.\ Phys.\ B {\bf 509} (1998) 33, hep-th/9707126.
%%CITATION = HEP-TH 9707126;%%
}
\lref\rfPeetersVW{K. Peeters, P. Vanhove and A. Westerberg, {\sl
Supersymmetric higher-derivative actions in ten and eleven dimensions, the
associated superalgebras and their formulation in superspace},
Class. Quant. Grav. {\bf 18} (2001) 843, hep-th/0010167.
%%CITATION = CQGRD,18,843;%%
}
\lref\rfGrossmanPLB{B. Grossman, T. Kephart and J.
Stasheff, {\sl Solutions To Gauge Field Equations In Eight-Dimensions:
Conformal Invariance And The Last Hopf Map}, Phys. Lett. {\bf B220} (1989)
431.
%%CITATION = PHLTA,B220,431;%%
}
\lref\rfGrossmanCMP{B. Grossman, T. Kephart and J.
Stasheff, {\sl Solutions To Yang-Mills Field Equations In Eight-Dimensions
And The Last Hopf Map}, Commun. Math. Phys. {\bf 96} (1984) 431.
%%CITATION = CMPHA,96,431;%%
}
\lref\rfTchrakian{D. Tchrakian,
{\sl Spherically Symmetric Gauge Field Configurations With Finite Action
In 4 P-Dimensions (P = Integer)}, Phys. Lett. B {\bf 150} (1985) 360.
}
\lref\rfDuffLuPRL{M.J. Duff and J.X. Lu, {\sl Strings from five-branes},
Phys. Rev. Lett. {\bf 66} (1991) 1402.
%%CITATION =     PRLTA,66,1402;%%
}
\lref\rfDuffLuCQG{M.J. Duff and J.X. Lu, {\sl A Duality between strings and
five-branes}, Class. Quant. Grav. {\bf 9} (1992) 1.
%%CITATION = CQGRD,9,1;%%
}
\lref\rfAbeSakai{N. Sakai and M. Abe, {\sl Coupling Constant Relations And
Effective Lagrangian In The Type I Superstring}, Prog. Theor. Phys.  {\bf 80}
(1988) 162.
%%CITATION = PTPKA,80,162;%%
}
\lref\CederwallTD{M.~Cederwall, B.~E.~Nilsson and D.~Tsimpis, {\sl D =
10 superYang-Mills at $O(\alpha^2)$}, hep-th/0104236.
%%CITATION = HEP-TH 0104236;%%
}
\lref\CederwallBT{M.~Cederwall, B.~E.~Nilsson and D.~Tsimpis, {\sl
The structure of maximally supersymmetric Yang-Mills theory:
Constraining higher-order corrections}, hep-th/0102009.
%%CITATION = HEP-TH 0102009;%%
}
\lref\BaggerWP{J.~Bagger and A.~Galperin, {\sl A new Goldstone multiplet for
partially broken supersymmetry}, Phys.\ Rev.\ D {\bf 55} (1997) 1091,
hep-th/9608177\semi
%%CITATION = HEP-TH 9608177;%%
T.~Hara and T.~Yoneya, {\sl Nonlinear supersymmetry
without the GSO projection and unstable  D9-brane}, Nucl.\ Phys.\ B {\bf 602},
 (2001) 499, hep-th/0010173\semi
%%CITATION = HEP-TH 0010173;%%
E.~Dudas and J.~Mourad, {\sl Consistent gravitino
couplings in non-supersymmetric strings}, hep-th/0012071\semi
%%CITATION = HEP-TH 0012071;%%
S.~Terashima and T.~Uesugi, {\sl On the
supersymmetry of non-BPS D-brane}, JHEP {\bf 0105}  (2001) 054,  hep-th/0104176.
%%CITATION = HEP-TH 0104176;%%
}
\lref\rfDundarer{
A. Dundarer, {\sl Multi-Instantons Solutions in
Eight-Dimensional Curved Space}, Mod. Phys. Lett. {\bf A6} (1991) 409.
}
\lref\obt{G. O'Brien and D. Tchrakian,
{\sl Spin Connection Generalized Yang-Mills Fields On Double Dual
Generalized Einstein-Cartan Backgrounds}, J.Math.Phys.  {\bf 29} (1988) 1212.
%%CITATION = JMAPA,29,1212;%%
}
\lref\chadu{J. M. Charap and M. J. Duff, {\sl Gravitational effects on
Yang-Mills topology}, Phys. Lett. {\bf B69} (1977) 445.
%%CITATION = PHLTA,B69,445;%%
}
\lref\PolchinskiDF{J.~Polchinski and E.~Witten, {\sl Evidence for Heterotic -
Type I String Duality}, Nucl.\ Phys.\ B {\bf 460}, 525 (1996), hep-th/9510169.
%%CITATION = HEP-TH 9510169;%%
}
%%%%%%%%%%%%%%%%%%%%%%%%%%%%%%%%%%%%%%%%%%%%%%%%%%%%%%%%%%%%%%%%%%%%%%%%%%%%%%
\Title{\vbox{\baselineskip12pt
\hbox{hep-th/0106096}
\hbox{CPHT-S028.0601}
\hbox{SPHT-T01/63}
\hbox{YCTP-SS6-01}
}}
{Closed Strings from SO(8) Yang-Mills Instantons}
\smallskip
\centerline{{\bf Ruben Minasian${}^1$, Samson L.
Shatashvili${}^2$\footnote{${}^*$}{\sevenrm On
leave of absence from St. Petersburg Steklov Mathematical
Institute, St. Petersburg, Russia.}} and
{\bf Pierre Vanhove${}^3$}}
\medskip
\centerline{${}^1$ \sl Centre de Physique Th{\'e}orique, {\'E}cole polytechnique,
91128 Palaiseau, France\footnote{${}^\circ$}{\sevenrm Unit{\'e} mixte du
CNRS
et de l'EP, UMR 7644.}}
\centerline{\tt ruben.minasian@cpht.polytechnique.fr}
\centerline{${}^2$ \sl Department of Physics, Yale University, New Haven, CT
06520-8120, USA}
\centerline{\tt samson.shatashvili@yale.edu}
\centerline{${}^3$ \sl Service de Physique Th{\'e}orique, CEA-Saclay,
F-91191 Gif-sur-Yvette Cedex, France.}
\centerline{\tt vanhove@spht.saclay.cea.fr}

\bigskip 

When eight-dimensional instantons, satisfying $F \wedge F = \pm \star_8 (
F\wedge F)$, shrink to zero size, we find stringy objects in higher order
ten-dimensional Yang-Mills (viewed as a low-energy limit of open string
theory). The associated $F^4$ action is a combination of two independent parts
having a single-trace and a double-trace structure.  As a result we get a
D-string from the single-trace term and a fundamental string from the
double-trace. The latter has $(8,0)$ supersymmetry on the world-sheet and
couplings to the background gauge fields of a heterotic string. A correlation
between the conformal factor of the instanton and the tachyon field is
conjectured.

\medskip

\Date{}%{\tt file:\jobname.tex} -- DO NOT DISTRIBUTE (YET!)}

%\vfill\supereject

%%%%%%%%%%%%%%%%%%%%%%%%%%%%%%%%%%%%%%%%%%%%%%%%%%%%%%%%%%%%%%%%%%%%%%%%%%%%%%%
\newsec{Introduction}

Recent developments in string theory have given some hope for answering
important questions which require off-shell formulation of a second
quantized string theory. In the case of open string field theory, several
off-shell formulations are  currently known \refs{\wone,
\ShatashviliKK,\BerkovitsBT}.  The situation with the closed string field
theory is more complicated. One might consider the possibility that in
fact the complete and consistent off-shell string theory can originate
from properly defined open string field theory alone, with closed string
being understood in terms of classical open string field theory (for
motivations and references see \refs{\ShatashviliUX, \GrossYK}).

The possibility that a closed string world-sheet action arises from classical
solutions in open string field theory will be the main theme of this paper.
Although a complete picture might require incorporating the whole field theory
of open strings, one might hope to get important insights by looking at low
energy sector only. This is what we will attempt here. We start from a
non-abelian super Yang--Mills theory including the higher-derivative $F^4$
terms in ten dimensions in the background of a $SO(8)$ instanton and find
stringy objects.\foot{The structure and the supersymmetry of $F^4$ terms will
be discussed in details in section 2.  For now we just remark that due to
different ways of contracting the gauge indices they can be of two
types---single-trace terms ${\rm tr}F^4$ and double-trace $\left({\rm
tr}F^2\right)^2$ terms.} The presence of higher-derivative terms in the
effective action is crucial for our analysis since not only the lowest order
$F^2$ term cannot lead to stringy objects with a finite tension but, as we
will argue, does not play any role.  In zero size limit of the instanton, a
D-string and a fundamental heterotic string are found. The heterotic string
solution will be supersymmetric only under a linear combination of the
standard supersymmetry of the Yang--Mills theory together with an extra
supersymmetry transformation. It is tempting to interpret this extra
supersymmetry transformation as originating from the non-linear supersymmetry
of a non-BPS D-brane sitting in our vacuum.

The consistency of the heterotic string background is related to the original
symmetries of the Yang--Mills background; the chirality of the background will
be reflected in the chirality of the string world-sheet theory.  In particular,
starting from an ``anomaly free" open string theory (a correct  choice
of the gauge group for which the top term in the anomaly polynomial
vanishes is what is meant here) will surely result in a closed heterotic 
string theory in a consistent gravitational and Yang--Mills background.

The D-string appears from the presence of ${\rm tr}F^4$ in the
effective action, and its tension is quantized according the value of
$\Pi_7(G)=\ZZ$ for $G$ allowing non-trivial embedding of $SO(8)$ groups
\refs{\rfDuffLuPRL,\rfWittenKtheory}. For obvious reasons, we chose 
$G=SO(32)$. The novelty of our study resides in the origin of a fundamental 
heterotic string from the double-trace structure $\left({\rm tr}F^2\right)^2$ 
and its supersymmetric completion in the effective action.

In the next section, the precise structure of these terms will be
shown from the supersymmetry analysis but we would like to make some
qualitative comments here from a different point of view.  Some
relevant ideas come just from looking at the (bosonic part of) open
string action 
\eqn\ostr{ S = S_{\rm open} (\phi, B, T) - \int
d^{10}x \sqrt{{\rm det}(G+F)} f(T) } 
written here for the constant tachyon field $T$ (the precise analog of
this action for non-abelian gauge groups is not known yet but can be
conceptually derived in any given string field theory). According to
the well-known conjectures of Sen \refs{\sen}, $f(T)$ has zero at the
minimum of the tachyon potential where the closed string vacuum is
expected. It can be calculated using the formalism of
\refs{\ShatashviliKK}; for bosonic string field theory it is given by
$f(T)=e^{-T}$ while the tachyon potential is $V_B=(1+T)e^{-T}$ and for
the superstring, $f(T)$ coincides with the tachyon potential
$V_F(T)=e^{-T^2}$ \refs{\GerasimovGA}. Since our analysis relies on
supersymmetry, from now on we will replace $f$ by $V$. 

In a general string field theory the lagrangian is defined up to field
redefinitions and a correct
choice of fields is dictated by the principle that equations of motion are
proportional to world-sheet $\beta$-function with some Zamolodchikov metric
$G$: $\partial_i S = G_{ij} \beta^j$. Here $G$ is the metric on the space of
fields and has to be non-degenerate.  One can note that the perfect square
part of double-trace terms, $\left({\rm tr}F_{\mu \nu} F^{\mu
\nu}\right)^2$, in a classical  open string field theory  lagrangian can
be generated by a field redefinition $T \rightarrow T+trF^2$. The
compatibility of the supersymmetry and the above principle should fix
these  terms unambiguously as well as complete the needed $t_{(8)}
\left({\rm tr}F^2\right)^2$.

We conclude this introduction by a comment on the world-sheet topology.
We already mentioned that by expanding around the solution we
obtain couplings for collective modes which turn out to reproduce
those of the heterotic string, and this should already indicate that the string
under the consideration is closed. Much of this structure is directly
inherited form the $SO(8)$ instanton construction. It is yet another
feature of the construction that the solution has in fact $SO(9)$ symmetry
and the fact that we are using only the $SO(8)$ part leads to 
the periodicity of one of the string world-sheet coordinates. 

A brief outline of the paper is the following. In section 2, we present the
Yang-Mills action and the supersymmetry transformations. In section 3, we look
for BPS configurations associated with eight-dimensional instanton
configurations having $SO(8)$ symmetry. We consider the supersymmetric
excitations in the section 4 and show the emergence of the two-dimensional
action with couplings of the heterotic string with a particular embedding of
the gauge group. Finally the section 5 contains a general discussion of our
results. We have collected various definitions and
identities needed in the main text  in two appendices.

%%%%%%%%%%%%%%%%%%%%%%%%%%%%%%%%%%%%%%%%%%%%%%%%%%%%%%%%%%%%%%%%%%%%%%%%%%%%%

\newsec{The effective action}

Since we need to  consider non-abelian and supersymmetric extensions of the
effective theory~\ostr,  we start by recalling some facts about
supersymmetrisation of open string effective actions.

%%%%%%%%%%%%%%%%%%%%%%%%%%%%%%%%%%%%%%%%%%%%%%%%%%%%%%%%%%%%%%%%%%%%%%%%%%%%%
\subsec{The linear supersymmetry}

The effective action can be expanded in power series of $\alpha'$, and
at each order supersymmetric action can be obtained by Noether
procedure. This construction is valid for any gauge group. This is done
in the
context of {\it linearly realized\/} ${\cal N}_{10}=1$ supersymmetry
transformations. The  lagrangian for the non-abelian degrees of freedom
constructed by this procedure up to and including $(\alpha')^2 F^4$ terms is
\refs{\CederwallTD}:

\eqn\eLnonAb{\eqalign{
{\cal L}=&-{1\over 4}F^A{}_{\mu\nu}F^A{}_{\mu\nu}- 8\bar\chi^A\slashed{D}\chi^A\cr
%&\left.\hphantom{-{1\over 4}F^A{}_{\mu\nu}F^A{}_{\mu\nu}}
&-{(\pi\alpha')^2\over 32} M_{ABCD}\left[(F^A{}_{\mu\nu}F^B{}_{\nu\mu})(
F^C{}_{\rho\sigma}F^D{}_{\sigma\rho})-4
F^A{}_{\mu\nu}F^B{}_{\nu\rho}F^C{}_{\rho\lambda}F^D{}_{\lambda\mu}\right]\cr
%\phantom{-6(\alpha')^2\Bigl\{}
&+(\pi\alpha')^2 M_{ABCD}\left[4F^A{}_{\mu\lambda} F^B{}_{\lambda\nu}(\bar\chi^C\Gamma_\mu (D_\nu\chi)^D)
+2F^A{}_{\mu\nu}F^B{}_{\rho\lambda}(\bar\chi^C\Gamma_{\mu\nu\rho}(D_\lambda\chi)^D)\right]\cr
&+ \hbox{\rm quartic fermions}\, .
}}
This action~\eLnonAb\ is invariant under the linear supersymmetry
transformations (by a straightforward generalization of the
results of  \refs{\rfPeetersVW})
\eqn\edeltaA{\eqalign{
\delta_\epsilon A_\mu ^A=& -4\bar\epsilon \Gamma_\mu \chi^A \cr
 - &\tfrac{(\pi\alpha')^2}{8} \, M^A{}_{BCD}\Big[ 2
(\bar\epsilon\Gamma_\mu\chi^B) F_{\nu\rho}^C F_{\rho\nu}^D - 8
(\bar\epsilon\Gamma^{\nu}\chi^B) F_{\nu\lambda}^C F_{\lambda\mu}^D
- (\bar\epsilon\Gamma^{\nu_1\cdots \nu_4}{}_\mu\chi^B)
F_{\nu_1\nu_2}^CF_{\nu_3\nu_4}^D\Big]
}}

\eqn\edeltachi{
\delta_\epsilon \chi ^A= {1\over 8}  \Gamma^{\mu\nu}\epsilon
F_{\mu\nu}^A
+\tfrac{(\pi\alpha')^2}{2^5\cdot 4!} M^A{}_{BCD} \Big[ \big(\teight{\nu}
\Gamma_{\nu_7\nu_8}\epsilon - \Gamma^{\nu_1\cdots \nu_6}\epsilon\big)
F_{\nu_1\nu_2}^BF_{\nu_3\nu_4}^C F_{\nu_5\nu_6}^D\Big]
}
$\Gamma^{\mu}$ are the $SO(1,9)$ $\Gamma$-matrices, as well as all the
spinors in this expression; the ten-dimensional space-time indices are
labeled by greek letters from the middle of the alphabet $\mu, \nu =
0,1,...,9.$

All the fields are in the adjoint of the gauge group $G=SO(32)$,
labeled by a capital letter from the beginning of latin alphabet. The quadratic
expressions in the field are contracted with the metric $\delta_{AB} =
{\rm tr}_{fund}(T_A T_B)$ and quartic expressions with the {\it completely
symmetric\/}  four rank tensor $M_{ABCD}$ \refs{\CederwallTD,\CederwallBT}.
For any group, there are {\it two independent\/} four-rank symmetric
tensors
given by the symmetrized trace in the fundamental representation ${\rm
tr}_{fund}(T_{(A} T_B T_C T_{D)})= {\rm Str}(\cdots)$ and the
double-trace $\delta_{(AB} \delta_{CD)}\sim {\rm
tr}_{fund}(\cdots){\rm tr}_{fund}(\cdots)$  in the fundamental representation
\eqn\eMabcd{
M_{ABCD} =m_1\,  {\rm tr}_{fund}(T_{(A} T_B T_C
T_{D)})+ m_2\,  \delta_{(AB} \delta_{CD)} \, .
}
The analysis of \refs{\CederwallTD,\CederwallBT} was purely in the
framework of ${\cal N}_{10}=1$ super-Yang-Mills theory without any reference
to a particular string vacuum. Linear supersymmetry fixes the independent
tensorials structures, or superinvariants, at each order in the
derivative expansion, up to a number of unfixed coefficients (not
affected by field redefinitions). We emphasize once more that in 
principle such structure can also be fixed from the requirement
that the equations of motion are proportional to the world-sheet
$\beta$-function.

When specifying the string vacuum of reference the coefficients multiplying
the group tensors become field-dependent functions. In particular these
coefficients will have a dependence on the
dilaton. The open string effective action is normalized as ${\cal S} = (\pi
\alpha')^2 \int d^{10}x\, {\cal L}/(g_s g_o^2)$ with the standard
normalisation value for the coupling constant $g^4_o = 2^{10}
\pi^7 (\alpha')^2 \, \kappa_{(10)}^2= 2^{16} \pi^{14} (\alpha')^6$
\refs{\rfAbeSakai}.  The first term in \eMabcd\ receives contributions
starting from the tree-level (disc) open string diagram: $m_1(\varphi)/g_s=
1/g_s +1 /2 + o(g_s)$; the second starts from one-loop open string diagrams:
$m_2(\varphi)/g_s=1/16+ o(g_s)$.  A dependence on the open string tachyon
field can appear too. As we will see this difference will lead eventually to
having D-string and a fundamental string tensions for solitons corresponding
to the single- and the double-trace parts respectively.  Looking at this
theory as the low-energy limit of the $SO(32)$ heterotic string, the
coefficients $m_1$ and $m_2$ will have different dilaton dependence. These two
different functional dependences on the dilaton have been shown
in~\refs{\BachasMC} to be compatible with the strong coupling duality between
type~I and heterotic string.  Bearing this in mind, we keep the field
dependence of the coefficients unspecified, relying only on their independence
under supersymmetrisation.

The presence of the two independent terms in~\eMabcd\ will be fundamental
for our analysis. The single-trace term will carry the charge of the
instanton introduced in section~3 and the double-trace term will be at the
origin of  the interactions of the fluctuations
around the instanton.

%%%%%%%%%%%%%%%%%%%%%%%%%%%%%%%%%%%%%%%%%%%%%%%%%%%%%%%%%%%%%%%%%%%%%%%%%%%%%
\subsec{The non-linear supersymmetry}

Looking at the effective action as related to D-brane configurations
in a type~II vacuum, it is natural to look for the possibility of
non-linear supersymmetry resulting from the breaking of some
of the original supersymmetries of the vacuum. There are various ways to
introduce non-linear supersymmetry  depending on
which fields $\Phi$ are coupled to the supersymmetry  parameter
\refs{\CederwallTD,\BaggerWP}
in $\delta_\eta \chi  = \eta + {\cal O}(\Phi) \eta$.

For the case of open string  actions with tachyon field and
abelian gauge field, the quadratic effective
action takes the form  (abelian field have tilde)
\eqn\eLAb{
{\rm L}={V(T)\over g_s}\left\{1-{1\over 4}\tilde F_{\mu\nu}\tilde F_{\mu\nu}
+\bar{\tilde \chi}\slashed{D}\tilde \chi +{1\over 2} \partial^\mu T \partial_\mu T +\cdots\right\}
}
and is invariant under the non-linear supersymmetry transformations
\eqn\eKappa{\eqalign{
\delta_\eta \tilde \chi = & \left(1 + {1\over \sqrt2} \partial^\mu T
\Gamma_\mu\right) \eta\cr
\delta_\eta T = & \sqrt2 \bar\eta \tilde \chi - 2 \partial^\mu T \bar\eta
\Gamma_\mu\tilde \chi - \sqrt2 \partial^\mu T \partial_\mu T \bar\eta\tilde\chi\cr
\delta_\eta \tilde A_\mu = & \bar\eta \Gamma_\mu\tilde\chi\,.
}}
For later reference we notice that the supersymmetry transformation on abelian
gaugino take the form $\delta_\eta \tilde \chi = g(T) \eta$ where $g(T)$ is
function of the tachyon field only, taken its values in the spinorial
representation of $SO(1,9)$.

In a way, the non-linear supersymmetry involving the tachyon field is
forced on us by the desire of working with pure Yang-Mills without the
presence of the supergravity multiplet. The open string vacuum, where this
second symmetry appears, will be important for us since as we will see in
section~4 a linear combination of the linear and non-linear
supersymmetries guarantees the supersymmetry of the fluctuations around
the instanton.

%%%%%%%%%%%%%%%%%%%%%%%%%%%%%%%%%%%%%%%%%%%%%%%%%%%%%%%%%%%%%%%%%%%%%%%%%%%%
\newsec{Introducing the SO(8) Instanton}

The main point of this paper is to search for closed strings
inside the field theory of open strings. From the point of view
of low-energy effective lagrangian this means that we shall look
for classical solutions in the sector of theory containing
the gauge fields and tachyon. Because we are looking for a string-like
solitonic objects it is natural to
investigate eight-dimensional gauge configurations. It is then not hard
to see that due to the gauge configuration having codimension two, an
expansion around such a soliton will result in a string. Understanding the
nature of this string  will be our task.

On a more technical level, the presence in the supersymmetry
transfomations of the gauginos  of terms linear and cubic in the the
gauge field strenghts is also indicative of a special role of
eight-dimensional gauge configurations. Luckily one such configuration,
namely the eight-dimensional $SO(8)$ instanton, is well known
\refs{\rfTchrakian,\rfGrossmanCMP}, and has been used for constructing
string solitons previously in \refs{\rfDuffLuPRL, \rfDuffLuCQG}.
We start from examining the $(\alpha')^2$ corrections to
$\delta_\epsilon\chi$ and observe that indeed it significantly simplifies
for the gauge field given by
\refs{\rfTchrakian,\rfGrossmanCMP}
\eqn\eSolution{
F^{ab}_{mn}(x) = 4 \left(\Gamma_{mn}{\cal P}_+\right)^{ab} \, {\rho^2\over (x^2 + \rho^2)^2},
}
where $\rho$ is the size of the instanton, $m,n=1,\cdots, 8$, and
$a,b$ are $SO(8)$ indices of positive chirality.\foot{We refer to appendix
A for details about $SO(8)$ instantons, in particular the chirality
discussion and their generalisations to multi-instantons solutions.}
Hereafter the form factor
will be abbreviated as $f(\rho, x) =
\rho^2/(x^2 + \rho^2)^2$.  We introduce the $SO(8)$ chirality projector
$$
{\cal P}_\pm = {1\over2} (1\pm \gamma_{(8)})\,.
$$
The $SO(1,9)$ fermionic indices in~\edeltaA\ and~\edeltachi\ are split
according to $SO(8)$ indices as
$\epsilon_\alpha=(\epsilon_a,\epsilon_{\tilde a})$ (note all the
ten-dimensional spinors are Majorana-Weyl). Most importantly for our
analysis the antisymmetric product of two
 $SO(1,9)$ $\Gamma$-matrices decompose under $SO(8)$ $\gamma$-matrices as
$$
\Gamma_{mn} = \gamma_{mn} {\cal P}_+ + \tilde \gamma_{mn} {\cal P}_-\, .
$$
As shown in appendix~A, the eight-dimensional solution~\eSolution\
satisfies a very important identity
\eqn\etheh{
M_{ABCD} \left(\thuit{r} + {1\over 2} \ehuit{r}\right) F_{r_1r_2}^A
F_{r_3r_4}^B F_{r_5r_6}^C= 0,
}
with the relative positive sign for our choice of the $SO(8)$
representation for the instanton. (For the single-trace part this has
been noticed in \rfDuffLuPRL.)

For the solution~\eSolution, using \etheh\ and the previous
conventions, the $(\alpha')^2$ correction to the gaugino supersymmetry
transformation~\edeltachi\ takes the form
\eqn\eFermionZM{
\delta^{(2)}_\epsilon\chi^A =
- {(\pi\alpha')^2\over 2^4\cdot 4!}\,
M^A{}_{BCD}\,\Big(\Gamma^{m_1\cdots m_6}{\cal P}_-\epsilon\Big) F_{m_1m_2}^BF_{m_3m_4}^CF_{m_5m_6}^D \, ,
}
the appearance of the projector ${\cal P}_-$ is related to the choice
of the $SO(8)$ representation used to define the instanton.

Since we have already restricted ourselves to eight-dimensional
configurations, it is easy to see that there may be further relations
between $\delta^{(0)}_\epsilon\chi$ and $\delta^{(2)}_\epsilon\chi$ by
virtue of Hodge duality. But in order to see these we need some facts
about eight-dimensional self-duality equations and their solutions
\refs{\rfTchrakian,\rfGrossmanCMP}, discussed in more details in
appendix A. The gauge field \eSolution\ is a solution of the following
self-duality equation in flat $\RR^8$ \refs{\rfTchrakian,\rfGrossmanCMP}
\eqn\insteq{M_{ABCD} \left(F^C\wedge F^D \mp \star_8 (F^C\wedge
F^D)\right) =
0\, .}
This equation is conformal, and the self-duality is guaranteed by the
properties of the $SO(8)$ $\gamma$-matrices.

One can check than the solution \eSolution\ also satisfies a relation
\eqn\eSDII{
{1 \over 2} m_1\left(4f(\rho,x)\right)^2\, \star_8 F^A = -\, M^A{}_{BCD} F^B\wedge F^C
\wedge F^D
}
Note that we are not using the curved (non-conformal) counterpart of
\insteq, but stay in the flat space and simply use
another property of the solution.
The equation \eSDII\ allows to establish a connection between the
quadratic and quartic term in the gauge field in~\eLnonAb\ and
$\delta^{(0)}_\epsilon\chi$ and $\delta^{(2)}_\epsilon\chi$.
Notice that in the lhs of \eSDII\ the dependence on $m_2$ has dropped out
by virtue of the first Pontryagin class $p_1(F)$ vanishing on the
solution. We finally rewrite the full resulting supersymmetry as
\eqn\edc{
\delta_\epsilon \chi ^A=  {1 \over 8} \Gamma^{rs} F_{rs}^A\left(
1+{4\over 3}(\pi\alpha')^2 m_1 (f(\rho,x))^2 {\cal P}_- \right) \epsilon\
\ .
}
Using~(B.7) to work out the contraction between the spacetime
$\Gamma$-matrices and the $SO(8)$ $\gamma$-matrices used to
construct the instanton we see that the supersymmetry transformation
of the $SO(8)$ part of the gaugino $\chi^{ab}$ takes the form
\eqn\eRed{
\delta_\epsilon \chi^{ab}_\alpha =-7\, f(\rho,x)\, ({\cal P}_+)^{ab}_{\alpha\beta} \left(1 + {4\over3}(\pi\alpha')^2 m_1(f(\rho,x))^2
{\cal P}_- \right)_{\beta\gamma} \epsilon_\gamma \, ,
}
that may be further reduced to\foot{The notation $({\cal
P}_+)^{ab}_{\alpha\beta}$
makes explicit that this operator projects on the positive $SO(8)$
chirality space labeled by the indices $a$ and $b$.}
\eqn\eReduction{
\delta_\epsilon \chi^{ab}_\alpha =
 -7\, f(\rho,x) \, ({\cal P}_+)^{ab}_{\alpha\beta} \epsilon_\beta\, ,
}
by remarking that the operator $({\cal
P_-})_{\alpha\beta}$ projects on the chirality space opposite to the
one used to construct the instanton
$$
({\cal P}_+)^{ab}_{\alpha\beta}{\cal P_-}_{\beta\gamma} = 0\, .
$$
Therefore  the
$(\alpha')^2$ contribution to the supersymmetry transformation
$\delta_\epsilon^{(2)}\chi$ vanishes identically. Since there is no
background for the fermions $\delta^{(2)}_\epsilon A_\mu=0$ for the
instanton~\eSolution\ and the $(\alpha')^2$ piece of the
lagrangian~\eLnonAb, in the $SO(8)$ bosonic background~\eSolution, is
invariant under the lowest-order linear supersymmetry transformations
\eqn\eInvariance{
\left.\delta_\epsilon^{(0)} {\cal L}^{(2)}\right|_{\rm for\ \eSolution} = 0\, .
}
From now on we will concentrate on this part of the Yang-Mills action.

The supersymmetry transformation~\eReduction\ is reminiscent of
the non-linear supersymmetry transformations~\eKappa\ $\delta_\eta
\tilde\chi = \eta$ up to the dependence on the profile of the
instanton. That the instanton~\eSolution\ realizes an identification
between the spacetime $SO(8)$ Lorentz group and the $SO(8)$ sub-group of
the gauge group, will be central, section~4, for getting a supersymmetric
sigma-model from the fluctuations around this instanton.

The effects of the instanton are all controlled by  the behavior of the form
factor $f(\rho,x)= \rho^2/(\rho^2+x^2)^2$. Running a little bit ahead of
the time one can already note that
for $\rho\to\infty$ or for $\rho\to0$ and not sitting on the instanton
$x\neq0$ the form factor goes to zero, and the limit corresponds to
the usual open string vacuum
\eqn\eOpen{
{\rm Open\ string\ vacuum:} \lim f(\rho,x)= 0\ {\rm for}\cases{ \rho\to\infty& for all $x$\cr \rho\to0&
and $x\neq0$} \, .
}
In this regime the linear supersymmetry~\edeltachi\ is not broken and
the non-linear supersymmetry~\eKappa\ is present. There is no relation between these two supersymmetries.

When sitting on the instanton $x=0$, the zero-size
limit behavior is different as the form factor blows up to infinity: this is
the closed string regime
\eqn\eClosed{
{\rm Closed\ string\ vacuum:}  \lim_{\rho\to0} f(\rho,0)\sim
\rho^{-2}=\infty\, .
}
In this regime the instanton dissolves and the non-linear
supersymmetry get promoted to linear supersymmetry  and a linear
combination of the two supersymmetries~\eReduction\ and~\eKappa\  is
expected to survive.

%%%%%%%%%%%%%%%%%%%%%%%%%%%%%%%%%%%%%%%%%%%%%%%%%%%%%%%%%%%%%%%%%%%%%%%%%%%%%
\newsec{The  string soliton}

Since for the instanton~\eSolution, the quadratic ${\cal L}^{(0)}$ and
quartic ${\cal L}^{(2)}$ pieces of the effective action~\eLnonAb\ are
decoupled (see~\eInvariance), and we can concentrate on the quartic
piece of the action showing how the sigma-model for an heterotic
string comes out
\eqn\eEA{
{\cal S}^{(2)} =-{(\pi\alpha')^2\over 2^9 \pi^7 (\alpha')^3g_s}\, \int d^{10}x\, {\cal
L}^{(2)}\, .
}
We shall now move to discussion of the string world-sheet
and  apply the standard technique of expansion around the soliton.
We take $x^m = x^m (\sigma^\alpha)$, where $\sigma^{\alpha}$ ($\alpha = 0,1$)
are two bosonic zero modes to be identified momentarily with the  
world-sheet coordinates, and  at lowest order in $z^m=x^m-x_0^m,$
\eqn\eDeformation{
\hat F_{\alpha m} = \partial_\alpha A_m =-4i \gamma_{mn} \left(\partial_\alpha
z^n f(\rho, x_0) + x^n (\partial_\alpha z \cdot x) f'(\rho,x_0) +
\cdots\right)\, .
}
As the size of the instanton goes to zero,  $\lim_{\rho\to0}(f(\rho,
x_0))^4=\delta^{(8)}(x_0)$  and this leads to emergence of
two-dimensional string world-sheet action from the corresponding
Yang-Mills action. Note that while the instanton solution 
enjoys a full $SO(9)$ \refs{\rfGrossmanCMP}, only the $SO(8)$
subgroup is left explicit by the identification of $\RR^8$ as the
complement of
the point $r_9=-\rho$. Any other point would have been as good. This
remaining rotational invariance reflects itself in the periodicity of the
world-sheet
coordinate $\sigma_1$. The construction puts no restriction on the
 world-sheet coordinate $\sigma_0$.
As far as the world-sheet action is concerned,
at the leading order only the first term matters and  realizes the
pull-back from the spacetime to the world-volume theory.
Using the decomposition of the antisymmetric product of two $SO(8)$
gamma-matrices as $(\gamma_{mn})^{ab} = \delta_{mn}^{ab} +
c_{mn}{}^{ab},$ where $c_{mnab}$ is the completely antisymmetric $SO(8)$
octonionic tensor, and covariantizing the
result one gets from the $F^4$ terms in~\eEA\ the expected
\eqn\strng{
{\cal S}^{(2)}= {1\over 2 \pi \alpha'} \int d^2 \sigma \left(2 {m_1\over g_s}
 + ( {17\over 1920} {m_1\over g_s} + {1\over 60} {m_2\over g_s})
 \partial_\alpha z_m  \partial^\alpha z^m + \cdots\right)
}
We would like to take this a bit further and in
particular study the coupling to background gauge fields.

Before we go on we comment on the coefficients in~\strng.  The first constant
term is the energy of the instanton, and thanks to the identities of
section~A.2, it receives contributions from the single-trace term only. It is
therefore proportional to $m_1$.  The second term from the fluctuations
around the instanton in the $SO(8)$ part of the gauge group receives
contributions from both the single- and the double-trace part, and depends on
$m_1$ and $m_2$.  We observe here that this leads to emergence of two
different string tensions here - the single-trace part
$m_1/(2\pi\alpha'g_s)\sim 1/(2\pi\alpha'g_s)$ has a tension of a D-string
while the double-trace part $m_2/(2\pi\alpha'g_s)\sim 1/(2\pi\alpha')$ has a
{\it fundamental string\/} tension.

A more interesting test comes from considering the next order perturbation
on the space of connections $A_m = A^{(0)}_m + A^{(1)}_m$
$(m=1,\cdots,8)$ around the instantonic solution~\eSolution\
$A^{(0)}_m$. The fluctuation $A^{(1)}_m = a_m$ lives in the $SO(24)$
part of the gauge group. Since we choose a purely bosonic instanton
background, the fermions are all of the first order type $\chi =
\chi^{(1)}$. Fermions, respectively in $SO(8)$ and $SO(24)$, are
associated with the fluctuations $\hat F_{\alpha m}$ and $A^{(1)}$.
At this point another crucial difference between the single- and double-trace
parts emerges. The former does not allow any mixing of the $SO(8)$ and
$SO(24)$ parts and an action \strng\ with half of the original supersymmetries
preserved is all one gets.  The double-trace term however contains the
supersymmetric extension
\eqn\eDeltaL{
\eqalign{
\Delta {\cal L}^{(2)} \sim  \Big[{}&
% {m_2\over 32\cdot 6}\, \thuit{\mu} {\rm tr}_{\rm
% SO(8)}(F_{\mu_1\mu_2} F_{\mu_3\mu_4})\, {\rm tr}_{\rm
% SO(8)}(F_{\mu_5\mu_6} F_{\mu_7 \mu_8})\cr
%+{}&
4m_2\,  {\rm tr}_{\rm SO(8)}(\bar\chi \Gamma_{\mu}
D_\nu\chi)\, {\rm tr}_{\rm SO(8)}(\hat F_{\mu\lambda} \hat F_{\lambda \nu})\cr
+{}&
2m_2\,  {\rm tr}_{\rm SO(8)}(\bar\chi
\Gamma_{\mu\nu\rho} D_\lambda\chi)\, {\rm tr}_{\rm
SO(8)}(\hat F_{\mu\nu} \hat F_{\rho\lambda})\Big]\, , \cr
}
}
whose supersymmetry variation under~\eReduction\ is
\eqn\eAddI{
\delta_\epsilon (\Delta {\cal L}^{(2)})
%\sim m_2 \, f(\rho,x)\times {\rm tr}_{\rm SO(8)}(\hat F^2) \,
%(\bar\epsilon \Gamma{\cal P}_+ D\chi )
\sim m_2 \, (f(\rho,x))^3\times (\bar\epsilon \Gamma{\cal P}_+ D\chi
) \,.
}
Contrary to the action evaluated for the classical background,
the action for the fluctuations is not invariant by itself under
the supersymmetry generated by $\epsilon$.
We have observed already the similarity between the linear and non-linear
supersymmetries, and it looks natural to try to use the latter in order to
compensate for $\delta_\epsilon (\Delta {\cal L}^{(2)})$.
As noticed below~\eKappa\ the geometry of the non-linear supersymmetry
transformation is controlled by the profile of the tachyon
field. Moreover, beyond the group theory identifications of the gauge
$SO(8)$ and the Lorentz $SO(8)$, a tachyon field dependence can occur in
the map between the two gaugini $\chi$ and $\tilde\chi$. We
confine this freedom to a function $W(T)$ whose relation
to the potential is yet unclear. Acting with the non-linear
supersymmetry~\eKappa\ on~\eDeltaL\ we get
\eqn\eAddII{
\delta_{\eta} (\Delta {\cal L}^{(2)}) \sim m_2 W(T)\times
% {\rm tr}_{\rm SO(8)}(F^2)\,
 (\bar\eta  \Gamma D\chi)\, ,
}
The fluctuations around the instanton~\eSolution\ can be made
supersymmetric under a linear combination of the two supersymmetry
transformations $\delta_\epsilon$ and $\delta_\eta$ provided
the following identification between the supersymmetry parameters
\eqn\eIdentification{
\eta = {(f(\rho,x))^3 \over W(T)} {\cal P}_+ \epsilon\, .
}
It follows that the action~\eDeltaL\ is supersymmetric granted
\eqn\eSusyOne{
 {\cal P}_- \eta=0
}
is satisfied and the positive-chirality component of  $\tilde\chi_+$
survives. Notice that this chirality is the one of the $SO(8)$
representation used to construct the instanton. The remaining
supersymmetry, generated by $\eta_+$, will give rise to light-cone
Green-Schwarz fermions for an heterotic string.

As we will see in the next subsection, these fermionic zero modes will
give rise to both the fermions of a chiral $(8,0)$ supersymmetric
sigma-model and the gauge degrees of freedom for the word-sheet theory
of an heterotic string obtained from the fluctuations around the
instantons~\eSolution.

%%%%%%%%%%%%%%%%%%%%%%%%%%%%%%%%%%%%%%%%%%%%%%%%%%%%%%%%%%%%%%%%%%%%%%%%%%%%

\subsec{Gauge coupling}

The coupling between $A^{(1)}_m = a_m$ to the world-sheet fermions
$\chi$, both in $SO(24)$, is essential for understanding the nature of
the string.  The effective action~\eLnonAb\ has only two terms with
ten-dimensional fermion bilinears, of the form $\chi \Gamma_{[1,3]}
D\chi F^2$. Once again, the gauge indices are contracted with the
tensor $M_{ABCD}$~\eMabcd\ containing a single-trace and a
double-trace structure. Since the single-trace part gives a
supersymmetric D-string, for which the $SO(24)$ is totally decoupled,
we concentrate here on the double-trace part only.

We proceed with decomposing the fermions into a $SO(8)$ part, $\xi$,  and
$SO(1,1)$ fermions
\eqn\eSplit{
\chi^A = \cases{ \xi(x) \otimes S(\sigma)& for $A\in{}$Adj(SO(8))\cr
\xi(x) \otimes \lambda^A(\sigma) & for $A\in{}$Adj(SO(24))\cr
}\, .}
$S^a$ is a eight-dimensional Majorana-Weyl fermion defined from the
SO(8) component of $\chi$ by $\chi^{ab}_\alpha= S_\beta \, ({\cal
P_+})^{ab}_{\alpha\beta}$. Its SO(8) chirality is the same as the one  used to construct the instanton~\eSolution.
The kinetic terms
for the fermions $S$ will come from the $SO(8)-SO(8)$ part of the action.
Since the fermions $\lambda$ are the fluctuations in the complement of the
$SO(8)$, used to construct the instanton, in $SO(32)$  
 the kinetic and gauge couplings
for the these come from the $SO(8)-SO(24)$ cross terms.

From the fermion bilinear terms in~\eLnonAb
$$
m_2\,\int d^{10}x \,  {\rm tr}_{\rm SO(8)}\left(\bar\chi\Gamma_\alpha
\partial_\beta\chi\right) {\rm tr}_{\rm SO(8)}(F_{\alpha n}F_{n \beta})
$$
we readily derive the
kinetic terms for the light-cone fermions
\eqn\eKineticS{
m_2\, \int d^2\sigma\,  \bar S \gamma_\alpha  \partial_\beta S\, h_{\alpha\beta}\, ,
}
with induced two-dimensional metric
$$
h_{\alpha\beta} :=\partial_\alpha z^n\partial_\beta z^n\times
\lim_{\rho\to0}\int d^8x\ (f(\rho,x))^2\ \bar\xi\xi(x)\, .
$$
Notice that~\eReduction\ implies that $\xi(x)$ scales as $f(\rho,x)$ and the
previous integral has a finite non-zero limit when $\rho$ goes to
zero. The other fermion bilinear in~\eLnonAb\ involving $\Gamma_{\mu\nu\rho}$
 does not  contribute to the kinetic term for the light-cone fermions $S$.

From the interactions between the $SO(8)$ deformation~\eDeformation\ and
the $SO(24)$ fermions
$$
m_2\,\int d^{10}x \,  {\rm tr}_{\rm SO(24)}\left(\bar\chi\Gamma_\alpha
\partial_\beta\chi\right) {\rm tr}_{\rm SO(8)}(F_{\alpha n}F_{n \beta})
$$
we derive the kinetic term for the twenty-four fermions $\lambda$
\eqn\eLambdaOne{
m_2\,\int d^2\sigma\, {\rm tr}_{\rm SO(24)}(\bar\lambda
\gamma_\alpha \partial_\beta\lambda)\, h_{\alpha\beta}\, ,
}
From the coupling
$$
m_2\,\int d^{10}x \,  {\rm tr}_{\rm SO(24)}\left(\bar\chi\Gamma_\alpha
[a_m,\chi]\right) {\rm tr}_{\rm SO(8)}(F_{\alpha n}F_{n m})
$$
we derive the coupling between the fermions $\lambda$ and the background
$SO(24)$ gauge field
\eqn\eLambdaTwo{
m_2\,\int d^2\sigma\, {\rm tr}_{\rm SO(24)}(\bar\lambda
\gamma_\alpha [a_m,\lambda])\, e_{\alpha}{}^m\, ,
}
where the contractions of the indices are performed with the vielbein
$$
e_\alpha{}^m~:= \partial_\alpha z^m \times \lim_{\rho\to0}\int d^8x\
(f(\rho,x))^2\ \bar\xi\xi(x)\, . 
$$
Finally the $\Gamma_{\mu\nu\rho}$ term in~\eLnonAb\ does not contribute.

Putting everything together, we arrive at the world-sheet theory for a
heterotic string in the light-cone gauge with the particular embedding of
the spin connection into the $SO(8)$ part of the $SO(32)$ gauge
group.\foot{Curiously enough, one of very few features of ordinary
four-dimensional instantons that is replicated by the $SO(8)$ instanton
is the possibility of extending the solution to a curved space by
setting the gauge connection equal to spin connection \refs{\obt}.}
In particular, the kinetic term proportional to $m_2$ in~\strng\
and~\eKineticS\ gives the $(8,0)$ supersymmetric sigma model invariant
under the supersymmetries generated by $\eta$. The kinetic term for the
$SO(24)$ fermions $\lambda$ is obtained in~\eLambdaOne\ which together
with the coupling~\eLambdaTwo\ to the external field, gives a
covariant derivative acting on the fermions.

It is well-known that the heterotic string can appear as a soliton in
type~I theory, and it is natural to inquire about the
difference of our mechanism from  the S-duality correspondance between
type~I and heterotic string of \refs{\PolchinskiDF}.  The latter requires
a background metric (with the Lorentz invariance broken to $SO(1,1) \times
SO(8)$) and a non vanishing vev for the dilaton, while we only have a
Yang-Mills background and a flat metric.

%%%%%%%%%%%%%%%%%%%%%%%%%%%%%%%%%%%%%%%%%%%%%%%%%%%%%%%%%%%%%%%%%%%%%%%%%%%%%%%
\newsec{Discussion}

We would like to underline some of the important points and
list some open questions. Starting from pure
ten-dimensional Yang-Mills action we are able to produce a D-string and
more surprisingly/controversially a heterotic string in the background of
eight-dimensional $SO(8)$ instanton. The former has a non-vanishing
vacuum  energy and supersymmetric $(8,0)$ world-sheet.
Its very plausible connections  to non-commutative solitons and to 
D1-D9 (D0-D8) systems are not very clear to us  at the moment and are
beyond the scope of this paper.  There may be some
interesting questions to be asked in this context.  As for the
heterotic string to which we will turn now, the $SO(9)$ symmetry of
the instanton solution gets reflected in the periodicty of one of the 
string world-sheet coordinates.

Our main interest is in the appearance of the fundamental {\it heterotic
string}. Technically speaking, its existence is due to the presence of the
double-trace structures in the Yang-Mills $F^4$ terms responsible for mixing
of $SO(8)$ and $SO(24)$. These terms were first seen from the supersymmetry
analysis on the ten-dimensional Yang-Mills action, and as we tried to argue
here may also be related to the field redefinition freedom (fixed by the
requirement that the equations of motion are proportional to the world-sheet
$\beta$-function and by supersymmetry) in the classical open string field
theory lagrangian. As a result by expanding around the soliton we obtain a
string action in the background where the $SO(8)$ gauge connection is set to
the spin connection.  It might be interesting to notice that this double-trace
structure also allows to generate kinetic terms for Yang-Mills fields in the
bulk by giving classical values to ${\rm tr}F^2$. Indeed, in the $SO(8)$
instanton background, $t_{(8)}\left({\rm tr}F^2 \right)^2 \rightarrow {\rm
tr}F_{\mu
\nu} F^{\mu \nu},$ and in a way the $SO(8)$ part is used as the ``center'' of
the gauge group.

Finally, a crucial point in the analysis was the mutual cancellation of
linear and non-linear supersymmetry. The latter is known for the abelian
case but applicable to the $SO(8)$ group by the magic of the $SO(8)$
instantons, which realizes an identification of the Lorentz and gauge
$SO(8)$ groups. The resulting connection between linear and non-linear
supersymmetry parameters contains a dependence of the tachyon and the size
of the instanton via $(f(\rho,x))^3/W(T)$. Of course it is natural to
expect that this ratio is a constant and thus one gets a correlation
between the tachyon and the size of the instanton (e.g. since $f(\rho,x)$
is singular when the instanton shrinks to the zero size, $W(T)$ should
behave accordingly). This correlation is not very explicit though, since
it is not yet completely clear how $W(T)$ is related to the tachyon
potential. One could hope to understand better what are the respective
regimes where the D-string and the heterotic string discussed above
dominate. Such a possibility of at least partially recovering the tachyon
potential from the low-energy analysis in our opinion deserves further
study.

\bigskip

{\bf Acknowledgements:} It is a pleasure to thank Carlo Angelantonj,
Costas Bachas, Emilian Dudas, Elias Kiritis and Hong Liu for
discussions, as well as Tigran Tchrakian for guidance into the $SO(8)$
instanton literature. The work of RM and PV is supported in part by
EEC contract HPRN-CT-2000-00122; RM is also supported by the INTAS
contract 55-1-590. The work of SS is supported in part by the OJI
award from DOE.

%%%%%%%%%%%%%%%%%%%%%%%%%%%%%%%%%%%%%%%%%%%%%%%%%%%%%%%%%%%%%%%%%%%%%%%%%%%%%
\vfill\supereject
\appendix{A}{Instantons in eight-dimensions}

In this appendix we review some facts about $SO(8)$ instantons, and
introduce the notations used in the main text. We also briefly discuss
different variants of the self-duality equations and their solutions.

\subsec{Definitions}

\item{\tria} The $SO(8)$ instantons satisfying the self-duality condition
in euclidean flat eight-dimensional $\RR^8$
\eqn\eSDI{
F_\pm\wedge F_\pm = \pm \star_8 (F_\pm\wedge F_\pm)= \pm {1\over 4!}
\varepsilon_{(8)} (F_\pm\wedge F_\pm)
}
have been constructed  in~\refs{\rfTchrakian,\rfGrossmanCMP}.
This solution  is defined on the
eight-sphere of radius $\rho$, $r_m r^m + r_9 r^9=\rho^2$  in
terms of projective coordinates  $(x_m)$ on $\RR^8$, identified with the
complement of the point $r_9=-\rho$
$$
r_m = \rho^2 \, {2 x_m\over \rho^2+x^2}\  (m=1,\cdots, 8), \qquad
r_9 = \rho {\rho^2-x^2\over \rho^2+x^2}\, .
$$
Thanks to the $SO(9)$ rotational invariance of the solution, we could
have considered the complement of $r_9=+\rho$, without any change in the
discussion of the main text. 
The solution with no singularity at the origin  is given by
\eqn\eA{
(A_\pm)_m = -4i\Gamma_{mn}{\cal P}_\pm\, {x^n \rho^2\over(\rho^2+x^2)^2}
}
and its curvature is
\eqn\eF{
(F_\pm)_{mn}= 4\Gamma_{mn}{\cal P}_\pm \, { \rho^2\over(\rho^2+x^2)^2}
}
where $\Gamma_{mn}{\cal P}_+=\gamma_{mn}$ belong to the ${\bf 8_s}$ 
representation of
$SO(8)$ and correspond to the ${}+{}$ sign in~\eSDI.
The other representation of opposite  $SO(8)$ chirality
$\Gamma_{mn}{\cal P}_-$ will be abbreviated as $\tilde \gamma_{mn}$.
The two representations correspond to the choice of sign in~\eSDI.
We refer to appendix~B for details on $\gamma$-matrix algebra.
These $\gamma_{mn}$ matrices are the $SO(8)$ equivalent of 't Hooft
eta-symbols and this solution generalizes the Jackiw-Rebbi instanton.

\noindent
 In the main text we make the choice of the positive $SO(8)$
chirality solution $F_+$, and drop the ${}+{}$ subscript.

\item{\tria} There  is another solution \refs{\rfGrossmanCMP} related
to the one above by a conformal transformation, with a singular
behavior at the origin
\eqn\eAII{
(A_\pm)_m = -4i\Gamma_{mn}{\cal P}_\pm \, {x^n \rho^2\over x^2(\rho^2+x^2)^2}}
and its associated curvature is
\eqn\eFII{
(F_\pm)_{mn} = 4\Gamma_{mn}{\cal P}_\pm \, { \rho^6\over x^4(\rho^2+x^2)^2}\, .
}

\item{\tria} The solution~\eF\ solves
\eqn\eSDII{
\star_8 F_\pm = \mp  F_\pm\wedge F_\pm\wedge F_\pm\, ,
}
with a conformal metric
\eqn\eMetricC{
g_{mn} = 4f(\rho,x) \eta_{mn}\, .
}
\eFII\ has a form factor singular at the origin and therefore does not
qualify for such non-conformal generalisation.

\item{\tria} The solution~\eF\ satisfies a relation
\eqn\eSDIII{
8\times \left(f(\rho,x)\right)^2\star_8 F_\pm = \mp  F_\pm\wedge F_\pm\wedge F_\pm\, ,}
with the {\it flat\/} metric $g_{mn} = \eta_{mn}$.
Some of the discussion in  the main text relies heavily on this relation.

\item{\tria} One can show that in $4d$ dimensions
the following double-duality relation holds for any Einstein metric
$$
(2d)!^2\,(\wedge^dR )^{m_1\cdots m_{2d} }{}_{\mu_1\cdots\mu_{2d} } =
\epsilon^{m_1\cdots m_{2d}}{}_{n_1\cdots n_{2d} }
(\wedge^dR)^{n_1\cdots n_{2d}}{}_{\nu_1\cdots\nu_{2d}}\epsilon^{\nu_1\cdots\nu_{2d}}{}_{\mu_1\cdots\mu_{2d}}
$$
where $\wedge^dR$ is the completely antisymmetrized product of
$4d$-dimensional Riemann curvatures. Similarly to four-dimensional case
\refs{\chadu}, for $d=2$ this double-duality implies a single duality relation
\eSDI\ for a gauge field $A$ \refs{\obt}, provided
\eqn\spinga{
A^{\pm}_{\mu} = - {1 \over 2} \omega_{\mu}^{mn} {\cal P}_{\pm}
\gamma_{mn}  \,~;
\,\,\,\,\,\,\,
F^{\pm}_{\mu \nu} = - {1 \over 2} R_{\mu \nu}^{mn} {\cal P}_{\pm}
\gamma_{mn} \,.
}
Thus the procedure of setting the gauge connection to spin
connection enables one to extend the construction of the instanton to
curved backgrounds.

\item{\tria} This setup is generalized to multiple instanton solutions
classified by  maps of the eight-sphere $S^8$ onto itself \refs{\rfDundarer}:
\eqn\eFn{\eqalign{
(F_\pm)_{mn} =& 2{\partial_n w_p\partial_m w_q\over (1+w_l w_l)^2}
\Gamma_{pq}{\cal P}_\pm\,\cr
 w_l \Gamma_l =& \left(x_8 \II +\sum_{m=1}^7 x_m \Gamma_m\right)^N\, .
}}
Such  instanton carries a topological charge  classified
by $\Pi_7(SO(8))=\ZZ \oplus \ZZ$.

\item{\tria} Finally we remark that the $SO(8)$ instantons can be embedded in
any $SO(n)$ group with $n\geq 8$  ($\Pi_7(SO(n))=\ZZ$ for $n>8$)
but not in $E_8\times E_8$ group, since
there is no non-trivial embedding of $SO(8)$ in $E_8$. This fact is easily
understood by noticing that $E_8$ has no independent quartic
invariants ${\rm tr}(F\wedge F\wedge F\wedge F) \propto {\rm
tr}(F\wedge F)\wedge {\rm tr}(F\wedge F)$, and that ${\rm tr}(F\wedge
F)=0$ for the both the solutions~\eF\ and~\eFII.

\item{\tria} Since all the $F^4$ action is expressed in term of
the completely symmetric tensor \refs{\CederwallTD,\CederwallBT} we
define the topological charge carried by the instanton as
$$
q_\pm~:= {1\over 4!} \int d^8x M_{ABCD} F_\pm^A\wedge F_\pm^B\wedge
F_\pm^C\wedge F_\pm^D\, ,
$$
which can be rewritten using first and second Pontrjagin classes
$$
q_\pm = {1\over 4!} \int d^8x \left(-4 m_1\, p_2(F) + 2(m_1+m_2)\,
p_1(F)\wedge p_1(F)\right)\, .
$$
Since for the instantons~\eF, \eFII\ and~\eFn\ $p_1(F)=0$, the charge carried by the
instanton is
\eqn\eChargeIII{
q_\pm = {m_1\over 4!} \int d^8x \, p_2(F)\, .
}
\eF\ and~\eFII\ have charge $q_\pm=\pm 2^8\pi^4$, and~\eFn\ has charge $q_\pm=\pm2^8\pi^4\times N$.

%%%%%%%%%%%%%%%%%%%%%%%%%%%%%%%%%%%%%%%%%%%%%%%%%%%%%%%%%%%%%%%%%%%%%%%%%%%%%
\subsec{Identities}

We show that the $SO(8)$ instanton~\eF\ satisfies  the identity
\eqn\eDD{
M_{ABCD} \left( \thuit{r} \pm {1\over 2}\ehuit{r}\right)(F_\pm)^A_{r_1r_2} (F_\pm)^B_{r_3r_4} (F_\pm)^C_{r_5r_6} N_{r_7r_8}^D=0
}
for any matrix $N$ in the adjoint of the group, generalising the
identity used in~\refs{\rfDuffLuPRL}.
The completely symmetric tensor $M_{ABCD}$ of equation~\eMabcd\ (defined
in~\refs{\CederwallTD}) has two parts---the symmetrized
single-trace and  the double-trace. It is therefore sufficient to
check that
$$\eqalign{
{\rm tr}\left[\thuit{r} (F_\pm)_{r_1r_2} (F_\pm)_{r_3r_4}
(F_\pm)_{r_5r_6}N_{r_7r_8}\right]=& 2^5 (7 d - 11) (d -
4)(f(\rho,x))^2\, {\rm tr}((F_\pm)_{r_7r_8}N_{r_7r_8})\cr
{\rm tr}\left[  \ehuit{r}
(F_\pm)_{r_1r_2} (F_\pm)_{r_3r_4}
(F_\pm)_{r_5r_6}N_{r_7r_8}\right]=&\mp 2^4\cdot6!\,
(f(\rho,x))^2\, {\rm tr}((F_\pm)_{r_7r_8} M_{r_7r_8})
}$$
therefore for $d=8$~\eDD\ is satisfied.
$$\eqalign{
\thuit{r} {\rm tr}\left[(F_\pm)_{r_1r_2} (F_\pm)_{r_3r_4}\right]
{\rm tr}\left[(F_\pm)_{r_5r_6}N_{r_7r_8}\right]=-&2^8(d-1)(d-8)
(f(\rho,x))^2\,  {\rm tr}((F_\pm)_{r_7r_8} M_{r_7r_8})\cr
{\rm tr}\left[\ehuit{r}
(F_\pm)^A_{r_1r_2} (F_\pm)^B_{r_3r_4} (F_\pm)_{r_5r_6}N_{r_7r_8}\right]=0&
}
$$
therefore for $d=8$~\eDD\ is satisfied.

%%%%%%%%%%%%%%%%%%%%%%%%%%%%%%%%%%%%%%%%%%%%%%%%%%%%%%%%%%%%%%%%%%%%%%%%%%%%
\subsec{An action for the Instanton}

This solution has been shown~\refs{\rfDuffLuPRL,\rfDuffLuCQG} to be of
importance for constructing the string as a soliton for the five-brane in
ten-dimension. It was there shown that the solution satisfies the
equation-of-motion \refs{\rfGrossmanPLB}

\eqn\eEOMI{
{\cal E}^{(0)}\left(A\right)^\mu= {1\over f(\rho,x)} \, D_\nu \left( f(\rho,x) F^{\mu\nu}\right)=0\, ,
}
and
\eqn\eEOMII{
{\cal E}^{(2)}\left(A\right)^\mu  =- {(\pi\alpha')^2\over
g^2_o}{e^{-\varphi}\over 3!\cdot 8 g_s}
M^A{}_{BCD}\left(D_\nu
(F^B)^\mu{}_\nu F^C_{\rho\sigma} F^D_{\sigma\rho} - {1\over 4} D_\nu
(F^B F^C F^D)^\mu{}_\nu\right)
}
derived from the effective action~\eLnonAb.

Thanks to the equality~\eDD, ${\cal E}^{(2)}\left(A\right)^\mu =0$ and the
Bianchi identity, the instanton~\eSDI\ is an extremum of the $(\alpha')^2$
part of the effective action, ${\cal L}^{(2)}$, but not a global extremum.

%%%%%%%%%%%%%%%%%%%%%%%%%%%%%%%%%%%%%%%%%%%%%%%%%%%%%%%%%%%%%%%%%%%%%%%%%%%%%%%

\appendix{B}{Spinology, $t_8$-ology and $\Gamma$-gymnastic}

\item{\tria} For a non-abelian gauge field
\eqn\eteight{\eqalign{
 t_{(8)}{\rm tr} F^4 =&\hphantom{{}+{}} 8 {\rm tr}(F_{\mu\nu} F_{\nu\rho}
F_{\rho\lambda} F_{\lambda\mu})+16 {\rm tr}(F_{\mu\nu}F_{\rho\lambda} F_{\nu\rho}
F_{\lambda\mu} )\cr
&-4{\rm tr}(F_{\mu\nu} F_{\mu\nu} F_{\rho\sigma}
F_{\rho\sigma}) -2 {\rm tr}(F_{\mu\nu} F_{\rho\sigma} F_{\mu\nu}
F_{\rho\sigma})\cr
  t_{(8)} {\rm tr}F^2\, {\rm tr}F^2  =& -2 {\rm tr}(F_{\mu\nu} F_{\nu\mu})
{\rm tr}(F_{\rho\sigma} F_{\sigma\rho}) +16 {\rm tr}(F_{\mu\nu} F_{\nu\rho})
{\rm tr}(F_{\rho\sigma} F_{\sigma\mu})\cr
&-4 {\rm tr}(F_{\mu\nu} F_{\rho\sigma}) {\rm tr}(F_{\nu\mu} F_{\sigma\rho}) +
8 {\rm tr}(F_{\mu\nu} F_{\rho\sigma}) {\rm tr}(F_{\nu\rho} F_{\sigma\mu})\cr
}}
which is easily seen to be given by the following symmetrized trace formula
\eqn\estr{
{\rm tr} t_{(8)} F^4 = 4! {\rm Str}\left( F_{\mu\nu} F_{\nu\rho}
F_{\rho\sigma} F_{\sigma\mu} - {1\over 4}(F_{\mu\nu} F_{\mu\nu})^2\right)
}

\item{\tria} The $SO(1,9)$ Gamma matrices $\Gamma_\mu$ are
decomposed into $SO(8)$ gamma matrices as
\eqn\eGammaG{
\Gamma_\mu = \pmatrix{ 0 & \gamma_\mu \cr\tilde\gamma_\mu & 0}\, .
}
$\gamma_\mu$ acts from the space of spinor of positive $SO(8)$
chirality into the space of negative $SO(8)$
chirality: $\gamma_\mu: S_+ \to S_-$. $\tilde\gamma_\mu$ does the
opposite. Antisymmetric products of  even numbers of Gamma matrices
respects the direct sum $S_+ \oplus S_-$. In particular
\eqn\eGG{
\Gamma_{\mu\nu} = \pmatrix{\gamma_{\mu\nu}&0\cr 0
&\tilde\gamma_{\mu\nu}}=\gamma_{\mu\nu} {\cal P}_+ +
\tilde\gamma_{\mu\nu} {\cal P}_-
}

\item{\tria} The antisymmetric product of two $SO(8)$ matrices can be
expressed in terms of the completely antisymmetric octonionic structure
constants

\eqn\eoct{
(\gamma_{mn})^{ab} = \delta^{ab}_{mn} + c_{mn}{}^{ab}\, .
}

\item{\tria} We need the following $\Gamma$ matrices identities (all
antisymmetrizations are with unit weight)
\eqn\eGammaI{
\{\Gamma_\mu ,\Gamma_\nu\} = 2\eta_{\mu\nu}
}
\eqn\eGammaII{
 \Gamma_{pq}\Gamma^{rs}= \Gamma_{pq}{}^{rs} + 4 \Gamma_{[p}{}^{[s}
\eta_{q]}{}^{r]}+ 2 \eta^{sr}_{pq}
}
\eqn\eGammaVII{
\Gamma_{\mu\lambda} \Gamma_{\lambda\nu} = (d-2) \Gamma_{\mu\nu} +
(d-1) \eta_{\mu\nu}
}
\eqn\eGammaIII{
{\rm tr} \left(\Gamma_{r_1\cdots r_n} \Gamma^{s_m\cdots s_1}\right)=2^4\times
n!\times \eta_{r_1\cdots r_n}^{s_1\cdots s_n}
}
\eqn\eGammaIV{
\varepsilon_{(8)}{}_{r_1\cdots r_8} = \gamma_{r_1\cdots r_8}\gamma_{(8)}
}
\eqn\eGammaV{
 \varepsilon_{(8)}{}^{r_1\cdots r_n}{}_{s_1\cdots s_{8-n}}
\gamma^{r_n\cdots r_1}=n!\times \gamma_{s_1\cdots s_{8-n}}\gamma_{(8)}
}

\listrefs

\bye